\documentclass[apj]{emulateapj}
\usepackage{rotating}

\usepackage{graphicx}
\usepackage{color}
\usepackage{t1enc}

\shorttitle{X-ray emission from S\,308}
\shortauthors{Toal\'a et al.}

\begin{document}

\title{X-ray emission from the Wolf-Rayet bubble
  S\,308$^{\star}$}\thanks{$^{\star}$Based on observations obtained
  with XMM-Newton, an ESA science mission with instruments and
  contributions directly funded by ESA Member States and NASA.}


\author{ J.A.\ Toal\'{a}$^{1,2,\dagger}$, M.A.\ Guerrero$^1$, Y.-H.\
  Chu$^3$, R.A.\ Gruendl$^3$, S.J.\ Arthur$^2$, R.C.\ Smith$^4$, and
  S.L.\ Snowden$^5$} 

\affil{$^{1}$ Instituto de Astrof\'\i sica de Andaluc\'\i a, IAA-CSIC,
  Glorieta de la Astronom\'\i a s/n, 18008 Granada, Spain}

\affil{ $^{2}$Centro de Radioastronom\'\i a y Astrof\'\i sica,
  Universidad Nacional Aut\'onoma de M\'exico, Campus Morelia,
  Apartado Postal 3-72, 58090, Morelia, Michoac\'an, M\'{e}xico}

\affil{$^{3}$ Department of Astronomy, University of Illinois, 1002
  West Green Street, Urbana, IL 61801, USA} 

\affil{$^4$ NOAO/CTIO, 950 N.\ Cherry Avenue, Tucson, AZ 85719, USA}

\affil{$^5$NASA Goddard Space Flight Center, Code 662, Greenbelt, MD
  20771, USA}

\email{$^{\dagger}$toala@iaa.es}

\begin{abstract}
  The Wolf-Rayet (WR) bubble S\,308 around the WR star HD\,50896 is
  one of the only two WR bubbles known to possess X-ray emission.  We
  present \textit{XMM-Newton} observations of three fields of this WR
  bubble that, in conjunction with an existing observation of its
  Northwest quadrant, map most of the nebula.  The X-ray emission from
  S\,308 displays a limb-brightened morphology, with a central cavity
  $\sim$22\arcmin\ in size and a shell thickness of $\sim$8\arcmin.
  This X-ray shell is confined by the optical shell of ionized
  material.  The spectrum is dominated by the He-like triplets of
  \ion{N}{6} at 0.43~keV and \ion{O}{7} at 0.57~keV, and declines
  towards high energies, with a faint tail up to 1~keV.  This spectrum
  can be described by a two-temperature optically thin plasma emission
  model ($T_1\sim$1.1$\times$10$^6$~K, $T_2\sim$13$\times$10$^6$~K),
  with a total X-ray luminosity $\sim2\times$10$^{33}$~erg~s$^{-1}$ at
  the assumed distance of 1.5~kpc.
\end{abstract}

\section{Introduction}

Wolf-Rayet\,(WR) bubbles are the final result of the evolution of the
circumstellar medium (CSM) of massive stars with initial masses $M
\gtrsim$35 $M_\odot$.  These stars exhibit high mass-loss rates
throughout their lives, peaking during their post-main-sequence
evolution that involves a Red or Yellow Supergiant (RSG or YSG) or
Luminous Blue Variable (LBV) stage (e.g.,
\citealp{2003A&A...404..975M}) during which the mass-loss rate can be
as high as $10^{-4}$--$10^{-3}\,M_\odot$ yr$^{-1}$
\citep{2000A&A...360..227N}, although the stellar wind velocity is low
($10$-$10^{2}$~km~s$^{-1}$).  The final WR stage is characterized by a
fast stellar wind ($v_{\infty}\gtrsim 10^{3}$~km~s$^{-1}$), which
sweeps up, shocks, and compresses the RSG/LBV material.  Thin-shell
and Rayleigh-Taylor instabilities lead to the corrugation and eventual
fragmentation of the swept-up shell
\citep{{1996A&A...305..229G},{1996A&A...316..133G},{2003ApJ...594..888F},{2006ApJ...638..262F},{2011ApJ...737..100T}}.
Clumpy WR wind-blown bubbles have been detected at optical wavelengths
around $\sim$10 WR stars in our Galaxy
\citep{{1983ApJS...53..937C},{2000AJ....120.2670G},{2010MNRAS.409.1429S}}.
Their optical emission is satisfactorily modeled as photoionized dense
clumps and shell material \citep{1993A&A...272..299E}.

X-ray emission has been detected so far in only two WR bubbles,
NGC\,6888 and S\,308
\citep{{1988Natur.332..518B},{1994A&A...286..219W},
  {1998LNP...506..425W},{1999A&A...343..599W},{2003ApJ...599.1189C},
  {2002A&A...391..287W},{2005ApJ...633..248W},{2011ApJ...728..135Z}}.
The most sensitive X-ray observations of a WR bubble are those of the
northwest (NW) quadrant of S\,308 presented by
\citet{2003ApJ...599.1189C}.  Their \emph{XMM-Newton} EPIC-pn X-ray
spectrum of S\,308 revealed very soft X-ray emission dominated by the
\ion{N}{6} He-like triplet at $\sim$0.43~keV and declining sharply
toward higher energies.  This spectrum was fit with a two-temperature
optically thin MEKAL plasma emission model, with a cold main component
at $kT_1 = 0.094$~keV (i.e., $T_\mathrm{X}\sim 1.1 \times 10^6$~K),
and a hot secondary component at $kT_2 \sim 0.7$~keV contributing
$\leq$6\% of the observed X-ray flux. The comparison of the X-ray and
optical H$\alpha$ and [\ion{O}{3}] images of S\,308 showed that the
X-ray emission is confined by the ionized shell.

In this paper, we present the analysis of three additional
\textit{XMM-Newton} observations of S\,308, which, in conjunction with
those of the NW quadrant presented by \citet{2003ApJ...599.1189C}, map
90\% of this WR bubble (see \S2). In \S3 and \S4 we discuss the
spatial distribution and spectral properties of the X-ray-emitting
plasma in S\,308, respectively. In \S5 we present our results of the
X-ray emission from the central star in the WR bubble. A discussion is
presented in \S6 and summary and conclusions in \S7.


\section{\textit{XMM-Newton} Observations}
\begin{figure*}[!htbp]
\includegraphics[width=1.0\linewidth]{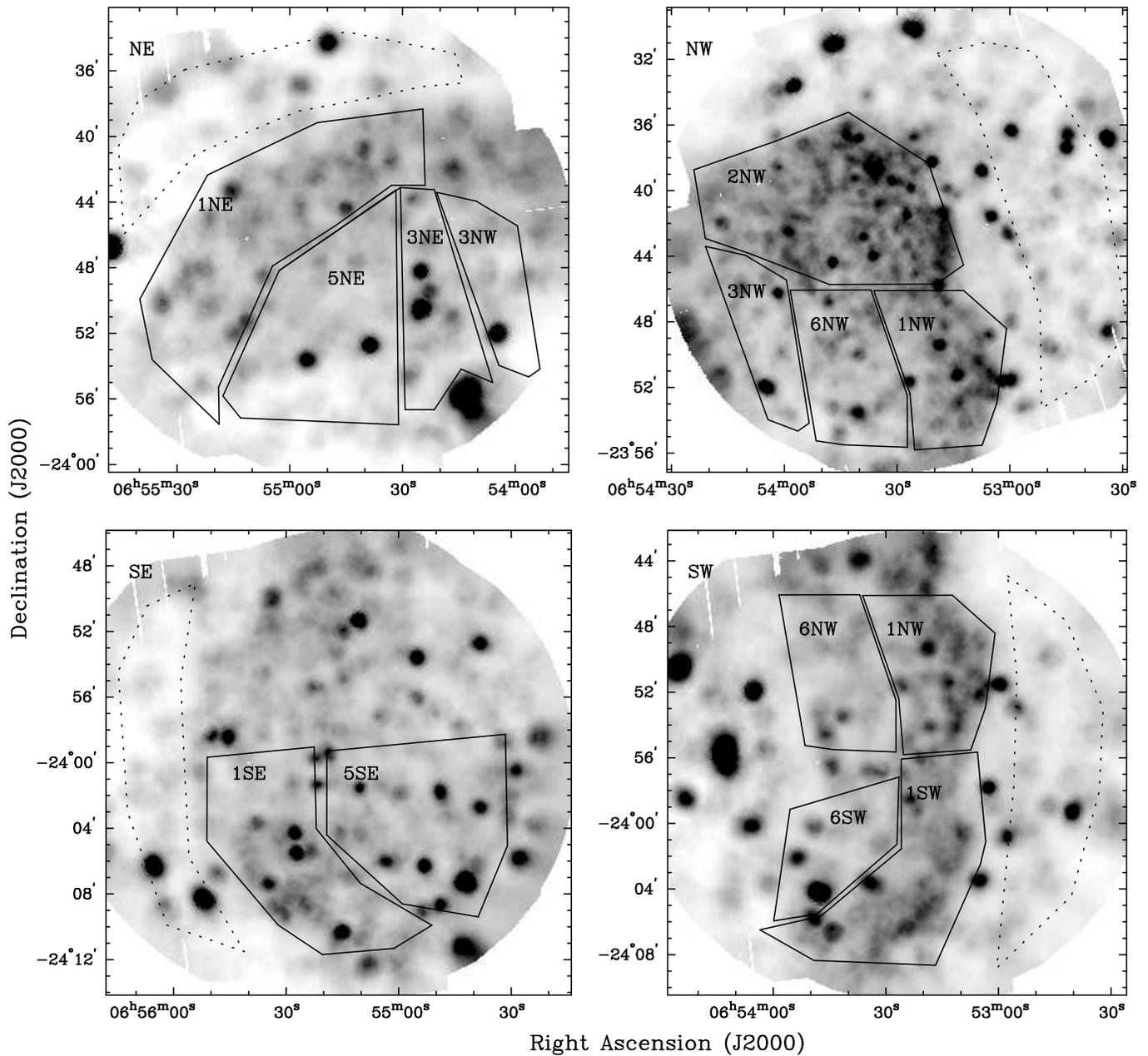}
\caption{\textit{XMM-Newton} EPIC images of the four observations of
  S\,308 in the 0.3--1.15~keV band.  The images have been extracted
  using a pixel size of 2\farcs0 and adaptively smoothed using a
  Gaussian kernel between 5\arcsec\ and 30\arcsec.  The source regions
  used for spectral analysis are indicated by solid lines and the
  background regions by dotted lines. Note that the point sources that
  are present in the images were excised for the spectral analysis.}
\label{fig:4images}
\end{figure*}

The unrivaled sensitivity of the \textit{XMM-Newton} EPIC cameras to
large-scale diffuse emission makes them the preferred choice for the
observation of S\,308.  \citet{2003ApJ...599.1189C} presented
\textit{XMM-Newton} observations of the brightest NW quadrant of
S\,308 \citep{1999A&A...343..599W}, but the large angular size of
S\,308 ($\sim$40\arcmin\ in diameter) exceeds the field of view of the
EPIC camera and a significant fraction of the nebula remained
unobserved.  To complement these observations, additional
\textit{XMM-Newton} observations of three overlapping fields covering
the northeast (NE), southwest (SW), and southeast (SE) quadrants of
the nebula have been obtained. The previous and new observations
result in the coverage of $\sim$90\% of the area of S\,308.
The pointings, dates and revolutions of the observations, and their
exposure times are listed in Table~\ref{tab:table1}.  In the
following, we will refer to the individual observations by the
quadrant of S\,308 that is covered, namely the NW, NE, SW, and SE
quadrants.  All observations were obtained using the Medium Filter and
the Extended Full-Frame Mode for EPIC-pn and Full-Frame Mode for
EPIC-MOS.

\begin{table*}[ht!]
\caption{\textit{XMM-Newton} Observations of S\,308}
\centering
\begin{tabular}{lcccccrrrcrrr}
\hline\hline\noalign{\smallskip}
\multicolumn{1}{l}{Pointing}            & 
\multicolumn{1}{c}{R.A.}                & 
\multicolumn{1}{c}{Dec.}                & 
\multicolumn{1}{c}{Obs.\ ID}            & 
\multicolumn{1}{c}{Rev.}                & 
\multicolumn{1}{c}{Observation start}   & 
\multicolumn{3}{c}{Total exposure time} & 
\multicolumn{1}{c}{}        &
\multicolumn{3}{c}{Net exposure time}   \\  
\cline{7-9}
\cline{11-13}
\multicolumn{1}{c}{}        &
\multicolumn{2}{c}{(J2000)} &
\multicolumn{1}{c}{}        &
\multicolumn{1}{c}{}        &
\multicolumn{1}{c}{UTC}     &
\multicolumn{1}{c}{pn}      &
\multicolumn{1}{c}{MOS1}    &
\multicolumn{1}{c}{MOS2}    &
\multicolumn{1}{c}{}        &
\multicolumn{1}{c}{pn}      &
\multicolumn{1}{c}{MOS1}    &
\multicolumn{1}{c}{MOS2}    \\
\multicolumn{1}{c}{}        &
\multicolumn{1}{c}{}        &
\multicolumn{1}{c}{}        &
\multicolumn{1}{c}{}        &
\multicolumn{1}{c}{}        &
\multicolumn{1}{c}{}        &
\multicolumn{1}{c}{[ks]}    &
\multicolumn{1}{c}{[ks]}    &
\multicolumn{1}{c}{[ks]}    &
\multicolumn{1}{c}{}        &
\multicolumn{1}{c}{[ks]}    &
\multicolumn{1}{c}{[ks]}    &
\multicolumn{1}{c}{[ks]}    \\
\hline
\noalign{\smallskip}
NW & 06:53:30 & $-$23:43:00 & 0079570201 & 343 & 2001-10-23T22:00:09 & 43.5 & 47.6 & 47.5 & & 11.9 & 19.6 & 19.9 \\
SW & 06:53:24 & $-$23:56:18 & 0204850401 & 781 & 2004-03-15T14:30:41 & 20.0 & 23.3 & 23.4 & &  6.4 & 9.0  & 9.2  \\	
SE & 06:55:16 & $-$24:00:00 & 0204850501 & 781 & 2004-03-14T23:00:41 & 22.0 & 25.4 & 25.4 & &  8.2 & 12.4 & 12.7 \\
NE & 06:54:47 & $-$23:46:18 & 0204850601 & 781 & 2004-03-15T06:45:41 & 22.0 & 25.4 & 25.4 & &  5.4 & 8.9  & 8.4  \\
\hline
\end{tabular}
\label{tab:table1}
\end{table*}

The \textit{XMM-Newton} pipeline products were processed using the
\textit{XMM-Newton} Science Analysis Software (SAS) Version 11.0, and
the Calibration Access Layer available on 2011-09-13.  In order to
analyze the diffuse and soft X-ray emission from S\,308, the
\textit{XMM-Newton} Extended Source Analysis Software (XMM-ESAS)
package \citep{2008A&A...478..615S,KS08} has been used.  This
procedure applies very restrictive selection criteria for the
screening of bad events registered during periods of high background
to ensure a reliable removal of the background and instrumental
contributions, particularly in the softest energy bands.  As a result,
the final net exposure times resulting from the use of the XMM-ESAS
tasks, as listed in Table~\ref{tab:table1}, are noticeably shorter
than the original exposure times.  Since we are interested in the best
time coverage of the central WR star to assess its possible X-ray
variability and given that its X-ray emission level is much brighter
than that of a mildly enhanced background, we applied less restrictive
criteria in selecting good time intervals for this star. For this
particular analysis, the 10--12~keV energy band is used to assess the
charged particle background, and we excised periods of high background
with EPIC-pn count rates $\geq$1.5~counts~s$^{-1}$ and EPIC-MOS count
rates $\geq$0.3~counts~s$^{-1}$.

\section{Spatial Distribution of the Diffuse X-ray Emission}

\subsection{Image processing}

Following Snowden \& Kuntz' cookbook for analysis procedures for
\emph{XMM-Newton} EPIC observations of extended objects and diffuse
background, Version 4.3 \citep{2011AAS...21734417S}, the XMM-ESAS
tasks and the associated Current Calibration Files (CCF), as obtained
from {\small
  \url{ftp://xmm.esac.esa.int/pub/ccf/constituents/extras/esas\_caldb}},
have been used to remove the contributions from the astrophysical
background, soft proton background, and solar wind charge-exchange
reactions, which have contributions at low energies ($<$1.5 keV).  The
resulting exposure-map-corrected, background-subtracted EPIC images of
each observed quadrant of S\,308 in the 0.3--1.15~keV band are
presented in Figure~\ref{fig:4images}.  The new observations of the
NE, SW, and SE quadrants of S\,308 detect diffuse emission, as well as
a significant number of point sources superimposed on this diffuse
emission.  With the single exception of HD\,50896 (a.k.a.\ WR\,6), the
WR star progenitor of this bubble registered in the SW and NE
observations, all point sources are either background or foreground
sources that we have removed prior to our analysis.

\begin{figure*}[!t]
\includegraphics[width=1.0\linewidth]{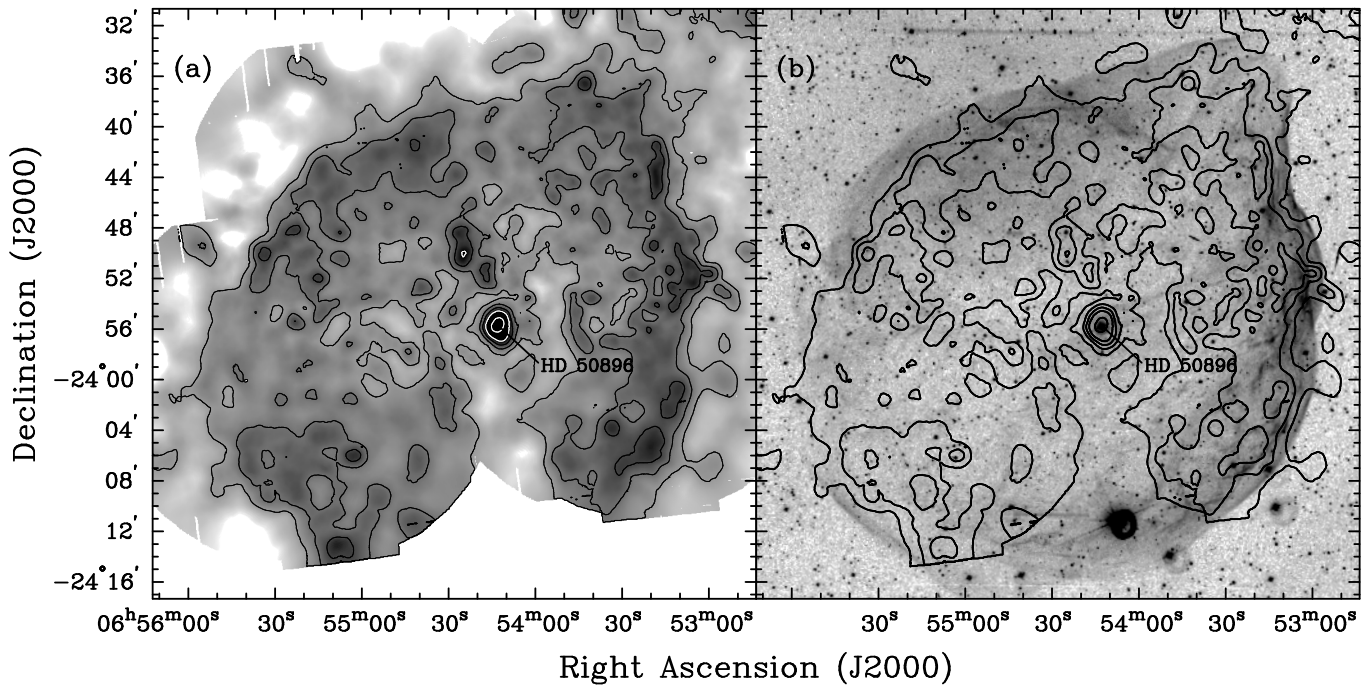}
\caption{
{\it (left)} Adaptively smoothed \emph{XMM-Newton} EPIC Image of S\,308 in
the 0.3--1.15~keV band.
All point sources, except for the central star HD\,50896 (WR\,6), have
been excised.
{\it (right)}
Ground-based [\protect\ion{O}{3}] image of S\,308 obtained with the Michigan
Curtis Schmidt telescope at Cerro Tololo Inter-American Observatory\,(CTIO) 
with superimposed X-ray emission contours.
The position of the central star HD\,50896 is indicated in both panels.
}
\label{fig:mosaic}
\end{figure*}

\begin{figure}[!t]
\includegraphics[bb=120 200 520 708,width=1.0\linewidth]{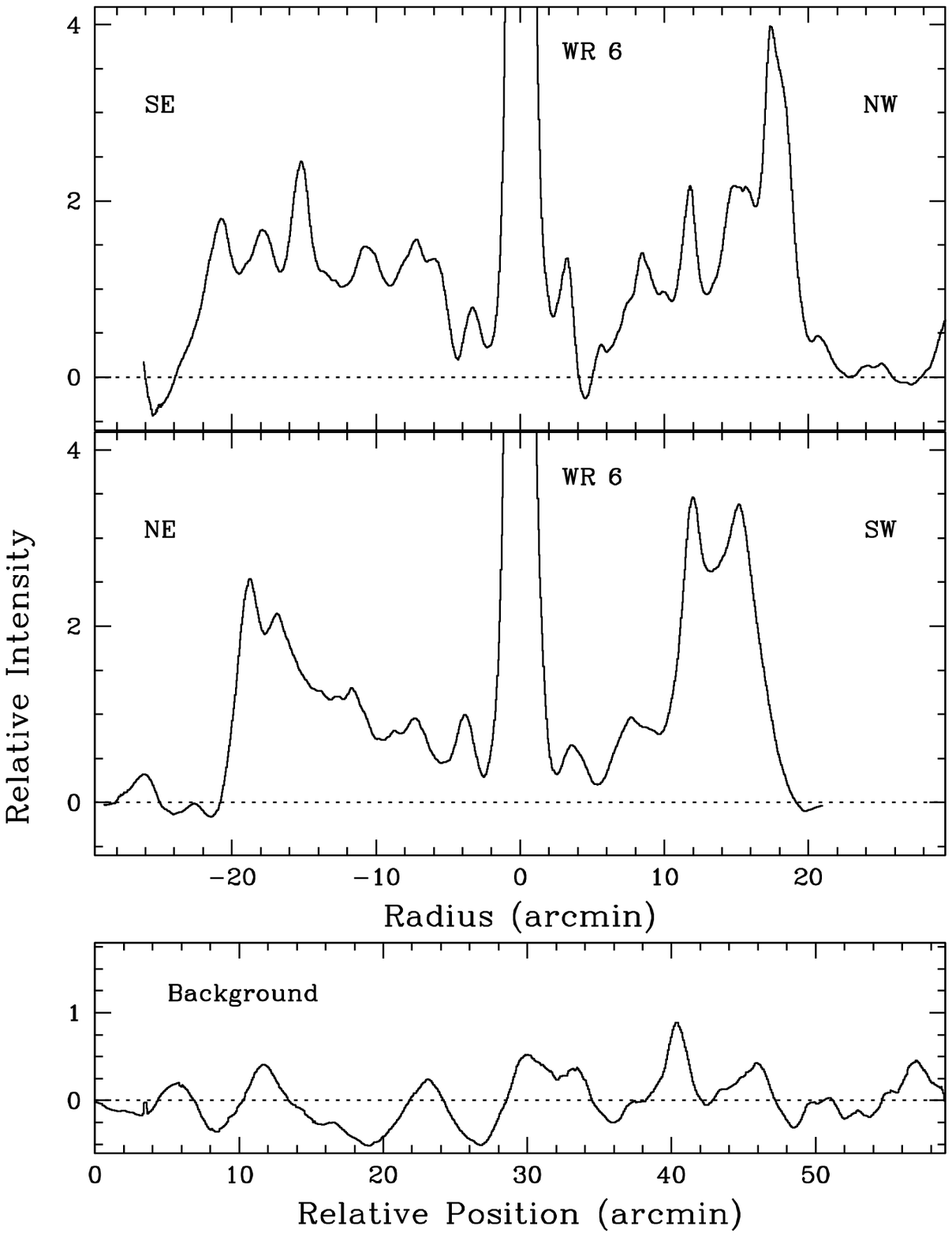}
\caption{ S\,308 X-ray surface brightness profiles along the SE--NW
  (PA=135$^\circ$) and SW--NE (PA=45$^\circ$) directions extracted
  from the smoothed \emph{XMM-Newton} EPIC image presented in
  Figure~\protect\ref{fig:mosaic}-\textit{left}.  For comparison, a
  surface brightness profile of a representative background region
  towards the West of S\,308 is shown at the same intensity and
    spatial scales.}
\label{fig:prof}
\end{figure}

\subsection{Analysis of the diffuse X-ray emission}

In order to analyze the spatial distribution of the diffuse X-ray
emission in S\,308, the four individual observations have been
mosaicked using the XMM-ESAS tasks and all point sources removed using
the \emph{Chandra} Interactive Analysis of Observations (CIAO) Version
4.3 \emph{dmfilth} routine, except the one corresponding to WR\,6.
The final image (Figure~\ref{fig:mosaic}-\textit{left}), extracted in
the 0.3--1.15~keV energy band with a pixel size of 3\farcs0, has been
adaptively smoothed using the ESAS task \emph{adapt-2000} requesting
100 counts of the original image for each smoothed pixel, with typical
smoothing kernel scales $\leqslant1'$ in the brightest regions and
$1'-2'$ in the faintest ones. This image reveals that the diffuse
X-ray emission from S\,308 has a limb-brightened morphology, with an
irregular inner cavity $\sim22'$ in size. The surface brightness
distribution displayed by this image confirms and adds further details
to the results of previous X-ray observations
\citep{1999A&A...343..599W,2003ApJ...599.1189C}.  The X-ray emission
from the bubble is brighter towards the northwest blowout and the
western rim, and fainter towards the east.  The bubble seems to lack
detectable X-ray emission towards the central regions around the WR
star.

\begin{figure*}[!t]
  \includegraphics[width=1.0\linewidth]{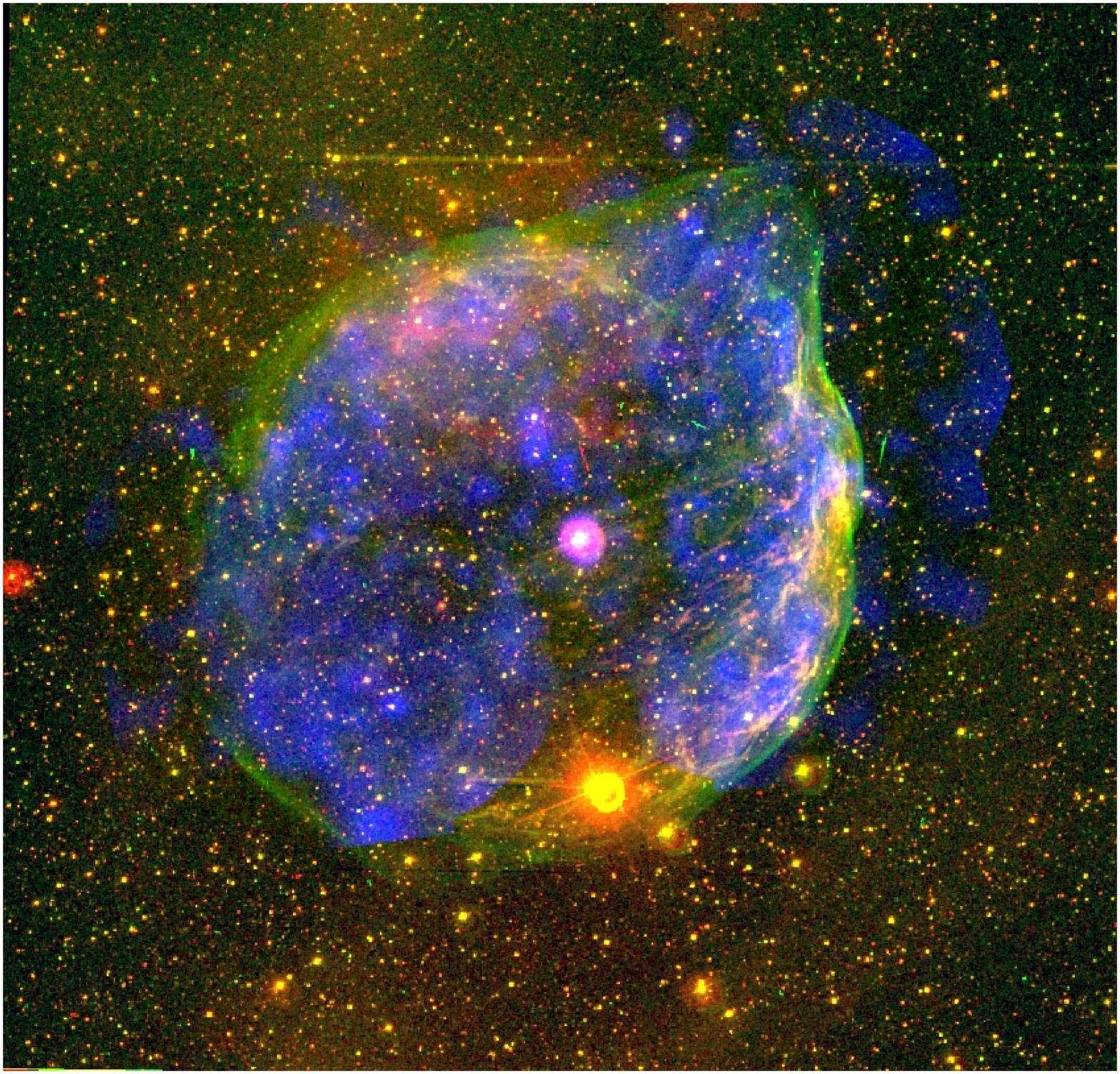}
\caption{Composite color picture of the \emph{XMM-Newton} EPIC image
  (blue) and CTIO [\ion{O}{3}] (green) and H$\alpha$ (red) images of
  S\,308. The apparent X-ray emission outside the optical shell is
  caused by large fluctuations in the background in regions near the
  EPIC cameras edge where the net exposure is much shorter than at the
  aimpoint.}
\label{fig:color_image}
\end{figure*}

The limb-brightened spatial distribution of the X-ray emission from
S\,308 is further illustrated by the surface brightness profiles along
the SE--NW and NE--SW directions shown in Figure~\ref{fig:prof}.  The
emission in the innermost regions, close to the central WR star, falls
to levels comparable to those of a background region to the west of
S\,308 shown in the bottom panel of Fig.~\ref{fig:prof}.  Besides the
SE region, whose surface brightness distribution is best described by
a plateau, the X-ray emission along the other directions increases
steadily with radial distance, peaking near the shell rim and
declining sharply outwards.  The thickness of the X-ray-emitting shell
is difficult to quantify; along the SW direction, it has a FWHM
$\sim$5\arcmin, whereas it has a FWHM $\sim$8\arcmin\ along the NE
direction.  Figure~\ref{fig:prof} also illustrates that the
X-ray-emitting shell is larger along the SE--NW direction
($\sim$44\arcmin\ in size) than along the NE--SW direction
($\sim$40\arcmin\ in size).

Finally, the spatial distribution of the diffuse X-ray emission from
S\,308 is compared to the [\ion{O}{3}] emission from the ionized 
optical shell in Figure~\ref{fig:mosaic}-\textit{right}.  The X-ray
emission is interior to the optical emission not only for the NW
quadrant but for the entire bubble.  This is also illustrated in the
color composite picture shown in Figure~\ref{fig:color_image}, in which 
the distribution of the X-ray emission is compared to the optical
H$\alpha$ and [\ion{O}{3}] images.  This image shows that the diffuse
X-ray emission is closely confined by the filamentary emission in the
H$\alpha$ line, whereas the smooth emission in the [O~{\sc iii}] line
extends beyond both the H$\alpha$ and X-ray rims.

\section{Physical Properties of the Hot Gas in S\,308}

The spectral properties of the diffuse X-ray emission from S\,308 can
be used to investigate the physical conditions and chemical abundances
of the hot gas inside this nebula.  In order to proceed with this analysis, 
we have defined several polygonal aperture regions, as shown in
Figure~\ref{fig:4images}, which correspond to distinct morphological
features of S\,308: regions with $\#$1 designations correspond to the 
rim revealed by the limb-brightened morphology, $\#$2 the NW blowout,
$\#$3, $\#$5, and $\#$6 shell interior, and $\#$4 the central star 
HD\,50896.  We note that any particular morphological feature may 
have been registered in more than one quadrant, in which case several 
spectra can have the same numerical designation (for instance, there 
are four spectra for the rim of the shell, namely 1NE,
1NW, 1SE, and 1SW).
\\
\\

\subsection{Spectra Extraction and Background Subtraction}

Perhaps the most challenging problem associated with the analysis of
the X-ray spectra of S\,308 is a reliable subtraction of the
background contribution.  The diffuse X-ray emission from S\,308 fills
a significant fraction of the field of view of the EPIC-pn and
EPIC-MOS cameras, making the selection of suitable background regions
difficult because the instrumental spectral response of the cameras
close to their edges may not be the same as those for the source
apertures.

The background contribution to the diffuse emission from clusters of
galaxies that fills the field of view is typically assessed from high
signal-to-noise ratio observations of blank fields. In the case of
S\,308, however, the comparison of spectra extracted from background
regions with those extracted from the same detector regions of the
most suitable EPIC Blank Sky observations \citep{2007A&A...464.1155C}
clearly indicates that they have different spectral shapes.  The
reason for this discrepancy lies in the typical high Galactic latitude
of the EPIC Blank Sky observations, implying low hydrogen absorption
column densities and Galactic background emission, while S\,308 is
located in regions close to the Galactic Plane where extinction and
background emission are significant.  We conclude that EPIC Blank Sky
observations, while suitable for the analysis of the diffuse emission
of a large variety of extragalactic objects, cannot be used in our
analysis of S\,308.

\begin{figure}[!t]
\includegraphics[bb=50 200 560 708,width=1.0\linewidth]{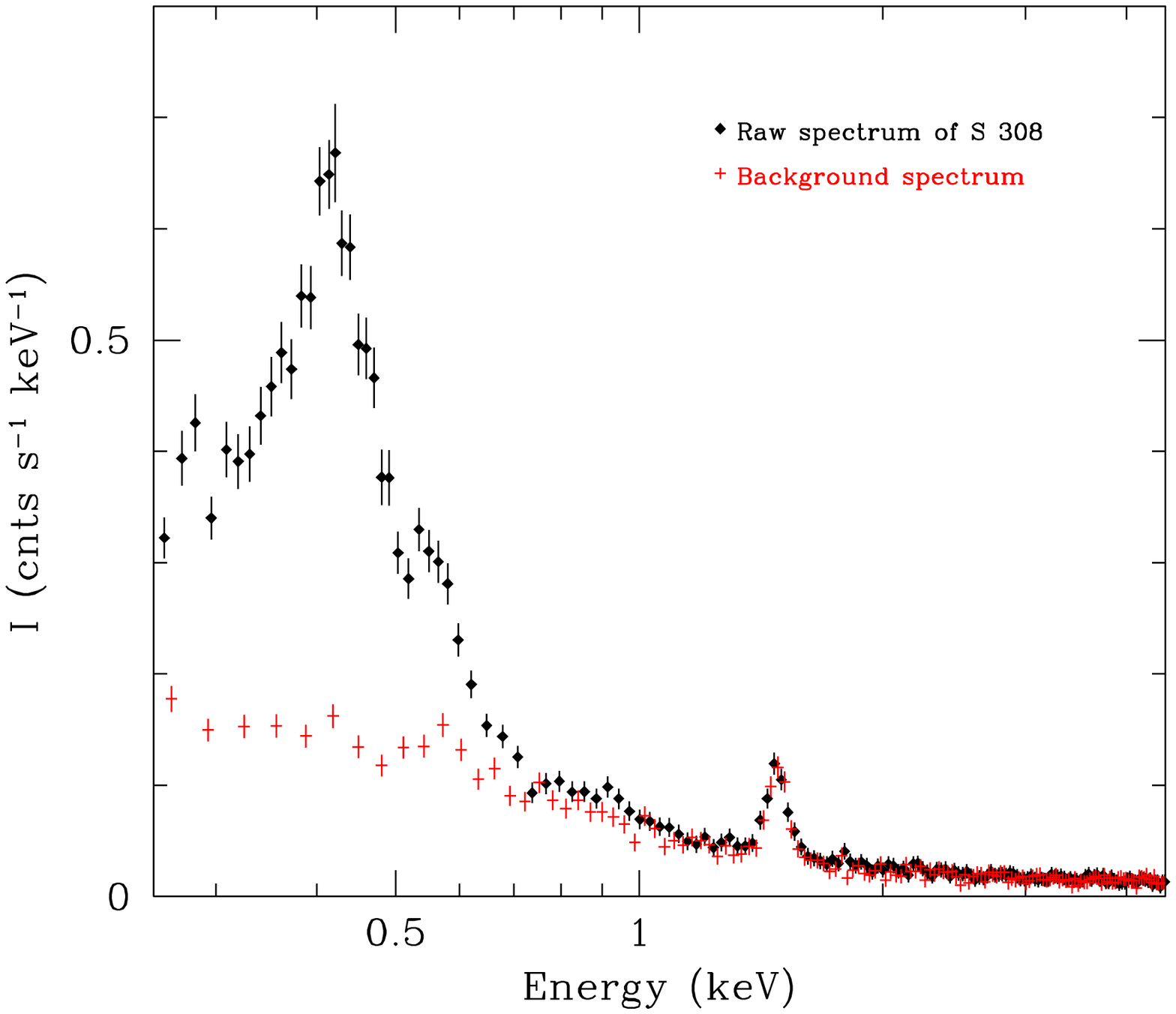}
\caption{Comparison of the background-unsubtracted raw EPIC-pn
  spectrum of S\,308 (black) and scaled EPIC-pn background spectrum
  (red). The Al-K line at $\sim$1.5 keV is an instrumental line.}
\label{fig:back}
\end{figure}

Alternatively, the different contributions to the complex background
emission in \emph{XMM-Newton} EPIC observations can be modeled, taking
into account the contributions from the astrophysical background,
solar wind charge-exchange reactions, high-energy (soft protons)
particle contributions, and electronic noise.  This is the procedure
recommended by the XMM-ESAS in the release of SAS v11.0, following the
background modeling methodology devised by \citet{Snowden_etal04} and
\citet{KS08}.  Even though the modeling of the different contributions
is a complex task, it can be routinely carried out.  Unfortunately,
S\,308 is projected close to the Galactic Plane and the \emph{ROSAT}
All Sky Survey (RASS) reveals that it is located in a region of strong
soft background emission with small-scale spatial variations. As shown
in Figure~\ref{fig:back}, the X-ray emission from this background is
soft and shows lines in the 0.3-1.0 keV energy band from thermal
components as the emission from S\,308.  Therefore, it is not possible
to model independently the emission from the S\,308 bubble and that
from the soft background.

\begin{figure}[!t]
\includegraphics[bb=48 175 548 710,width=1.0\linewidth]{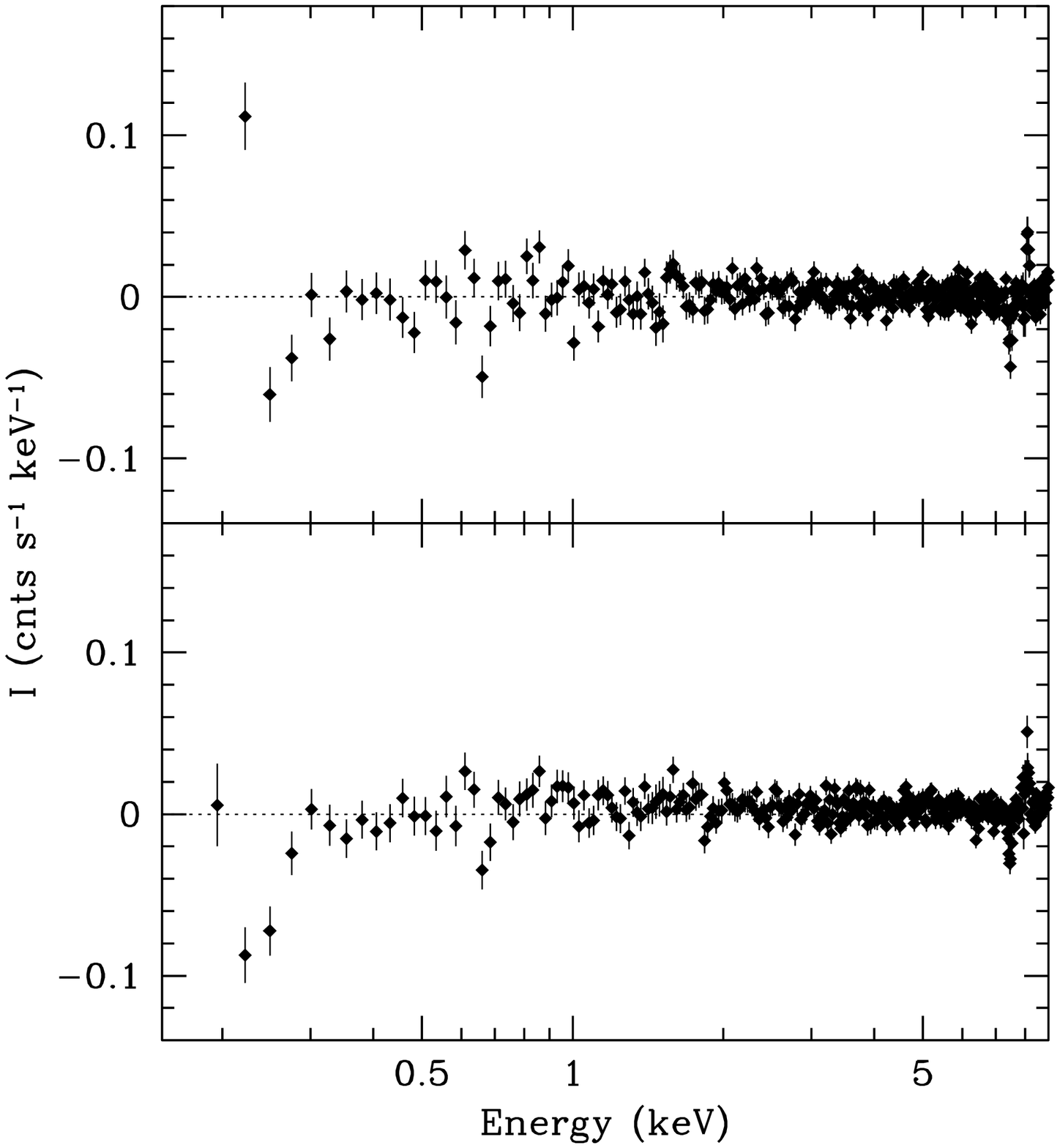}
\vspace{0.1cm}
\caption{
Background-subtracted blank sky spectra extracted from source and
background regions equal in detector coordinates to the regions 1NW
(\textit{top}) and 1SW (\textit{bottom}) of S\,308.
}
\label{fig:blank}
\end{figure}

The only viable procedure for the analysis seems the use of background
spectra extracted from areas near the camera edges of the same
observations.  It can be expected that the spectral properties of
these background spectra differ from those of the background
registered by the central regions of the cameras, given the varying
spectral responses of the peripheral and central regions of the
cameras to the various background components. To assess these
differences, we have used EPIC Blank Sky observations to extract
spectra from source and background regions identical in detector units
to those used for S\,308.  Two typical examples of blank sky
background-subtracted spectra are presented in Figure~\ref{fig:blank}.
While these spectra are expected to be flat, several deviations can be
noticed: (1) clear residuals at $\sim 7.5$ and $\sim 8.1$~keV, which
can be attributed to the defective removal of the strong instrumental
Cu lines that affect the EPIC-pn spectra \citep{KS08}, (2) a
noticeable deviation at $\sim0.65$~keV of the O-K line, and (3) most
notably, deviations at energies below 0.3~keV, which is indicative of
the faulty removal of the electronic noise component of the background
\citep{2007A&A...464.1155C}.  Note that the Al-K line at $\sim1.5$~keV
and the Si-K line at $\sim1.8$~keV, which may be expected to be strong
in the background EPIC-pn spectra, are correctly removed.
Consequently, we have chosen to use the background spectra extracted
from the observations of S\,308, but restrict our spectral fits to the
0.3--1.3~keV band.

\begin{figure*}[!htbp]
\includegraphics[bb=18 165 592 718, width=\linewidth]{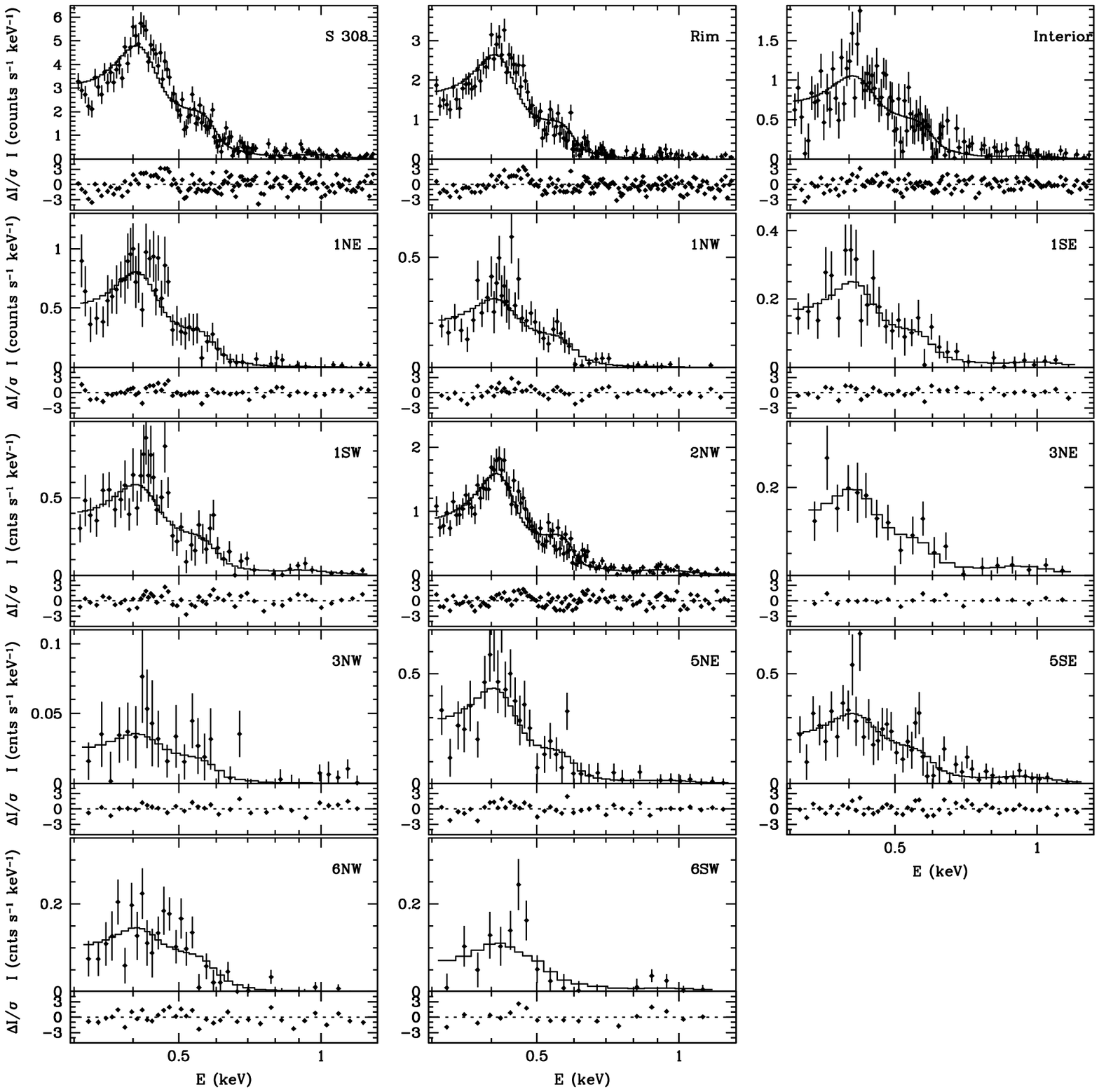}
\vspace{0.1cm}
\caption{Background-subtracted \emph{XMM-Newton} EPIC-pn spectra of
S\,308 corresponding to the 11 individual source regions shown in
Figure~\protect\ref{fig:4images}, as well as the combined spectra
of the entire nebula, its shell or rim (regions $\#1$ and $\#2$) and the 
interior region of the shell ($\#3$, $\#5$, and $\#6$).  Each spectrum is
overplotted with its best-fit two-temperature APEC model in the
energy range 0.3--1.3 keV, assuming fixed abundances, while the lower
panel displays the residuals of the fit.
}
\label{fig:spec_diffuse}
\end{figure*}

\subsection{Spectral properties}
\label{sec:specprops}

The individual background-subtracted EPIC-pn spectra of the diffuse
emission of S\,308 are presented in Figure~\ref{fig:spec_diffuse}.
This figure also includes spectra of the whole nebula, its rim, and
central cavity obtained by combining all spectra, those of regions
$\#$1 and $\#$2 (the rim or limb-brightened shell), and those of
regions $\#$3, $\#$5, and $\#$6 (the central cavity), respectively.
Spectra and response and ancillary calibrations matrices from
different observations of the same spatial regions were merged using
the \textit{mathpha}, \textit{addarf}, and \textit{addrmf} HEASOFT
tasks according to the prescriptions of the SAS threads. The EPIC-MOS
spectra have also been examined and found to be consistent with the
EPIC-pn ones although, due to the lower sensitivity of the EPIC-MOS
cameras, they have fewer counts.  Therefore, our spectral analysis of
the diffuse X-ray emission will concentrate on the EPIC-pn spectra.

The spectra shown in Figure~\ref{fig:spec_diffuse} are all soft, with
a prominent peak near $\gtrsim 0.4$~keV, and a rapid decline in
emission towards higher energies. Some spectra (e.g., 1NE, 1NW, 1SW,
5NE, and 5SE) show a secondary peak near $\lesssim 0.6$~keV that is
only hinted in other spectra.  There is little emission above $\sim
0.7$~keV, although some spectra (e.g., 1SE, 1SW, 2NW, 3NE, 5SE, and
6SW) appear to present a hard component between 0.8 and 1.0~keV.

\begin{table*}[ht!]
\caption{Spectral Fits of the Diffuse X-ray Emission of S\,308}
\scriptsize
\centering
\begin{tabular}{lrllcclccll}
\hline\hline\noalign{\smallskip}
\multicolumn{1}{l}{Region}        &
\multicolumn{1}{l}{Counts}        &
\multicolumn{1}{c}{$N_\mathrm{H}$}    &
\multicolumn{1}{c}{$kT_1$}        &
\multicolumn{1}{c}{EM$_1^{\mathrm{a}}$}         &
\multicolumn{1}{c}{$f_1^{\mathrm{b}}$}         &
\multicolumn{1}{c}{$kT_2$}        &
\multicolumn{1}{c}{EM$_2^{\mathrm{a}}$}         &
\multicolumn{1}{c}{$f_2^{\mathrm{b}}$}         &
\multicolumn{1}{c}{$f_2/f_1$}     &
\multicolumn{1}{c}{$\chi^2$/DoF}  \\
\multicolumn{1}{c}{}        &
\multicolumn{1}{c}{}        &
\multicolumn{1}{c}{[$10^{20}$cm$^{-2}$]}     &
\multicolumn{1}{c}{[keV]}                    &
\multicolumn{1}{c}{[cm$^{-3}$]}              &
\multicolumn{1}{c}{[erg\,cm$^{-2}$s$^{-1}$]} &
\multicolumn{1}{c}{[keV]}                    &
\multicolumn{1}{c}{[cm$^{-3}$]}              &
\multicolumn{1}{c}{[erg\,cm$^{-2}$s$^{-1}$]} &
\multicolumn{1}{c}{}                         &
\multicolumn{1}{c}{}                         \\
\hline
\noalign{\smallskip}
S\,308   & 10290$\pm$19 &~~~6.2   & 0.096$\pm$0.002 & 7.6$\times$10$^{56}$ & 2.7$\times$10$^{-12}$ & 1.12$\pm$0.22 & 1.2$\times$10$^{55}$ & 2.4$\times$10$^{-13}$ & 0.090       & 2.01 (=319.6/159) \\
Shell    & 7920$\pm$23  &~~~6.2   & 0.092$\pm$0.003 & 4.9$\times$10$^{56}$ & 1.5$\times$10$^{-12}$ &       $\dots$ & $\dots$             & $\dots$              & $\dots$      & 1.54 (=233.9/151) \\
Interior & 2390$\pm$120 &~~~6.2   & 0.116$\pm$0.011 & 8.3$\times$10$^{55}$ & 5.6$\times$10$^{-13}$ & 0.95          & 4.8$\times$10$^{54}$ & 7.5$\times$10$^{-14}$ & 0.134       & 1.35 (=188.2/139) \\
\hline
1NE    & 965$\pm$10  &~~~6.2     & 0.094$\pm$0.010 & 1.7$\times$10$^{56}$ & 5.1$\times$10$^{-13}$ & 0.95          & 1.1$\times$10$^{53}$ & 2.5$\times$10$^{-15}$ & 0.005          & 0.82 (=46.9/57) \\
1NW    & 1000$\pm$60 &~~~6.2     & 0.102$\pm$0.009 & 4.7$\times$10$^{55}$ & 2.0$\times$10$^{-13}$ & 0.95          & 4.4$\times$10$^{50}$ & 1.0$\times$10$^{-17}$ & $<10^{-3}$     & 1.13 (=57.9/51) \\
1SE    & 540$\pm$50  &~~~6.2     & 0.094$\pm$0.010 & 1.8$\times$10$^{56}$ & 5.1$\times$10$^{-13}$ & 0.95          & 1.1$\times$10$^{53}$ & 2.5$\times$10$^{-15}$ & 0.005          & 0.82 (=46.9/57) \\
1SW    & 1100$\pm$50 &~~~6.2     & 0.097$\pm$0.012 & 1.0$\times$10$^{56}$ & 3.9$\times$10$^{-13}$ & 0.95          & 2.3$\times$10$^{54}$ & 5.2$\times$10$^{-14}$ & 0.135          & 1.31 (=69.3/53) \\
\hline
2NW    & 4620$\pm$100 &~~~6.2    & 0.095$\pm$0.003  & 2.2$\times$10$^{56}$ & 7.4$\times$10$^{-13}$ &0.96$\pm$0.21  & 4.3$\times$10$^{54}$& 9.8$\times$10$^{-14}$ & 0.130         & 1.30 (=137.9/106) \\
\hline
3NE    & 300$\pm$33  &~~~6.2    & 0.095$\pm$0.023 & 3.5$\times$10$^{55}$ & 1.2$\times$10$^{-13}$ & 0.95           & 1.4$\times$10$^{54}$ & 3.1$\times$10$^{-14}$ & 0.262          & 0.55 (=9.4/17) \\
3NW    & 160$\pm$25  &~~~6.2    & 0.11$\pm$0.04   & 5.0$\times$10$^{54}$ & 2.5$\times$10$^{-14}$ & 0.95           & 5.6$\times$10$^{51}$ & 1.2$\times$10$^{-16}$ & 0.005          & 0.91 (=27.4/30) \\
\hline
5NE    & 530$\pm$50  &~~~6.2    & 0.090$\pm$0.015 & 7.6$\times$10$^{55}$ & 2.1$\times$10$^{-13}$ & 0.95           & 9.0$\times$10$^{53}$ & 1.8$\times$10$^{-14}$ & 0.083          & 1.06 (=44.5/42) \\
5SE    & 820$\pm$60  &~~~6.2    & 0.103$\pm$0.016 & 4.4$\times$10$^{55}$ & 1.9$\times$10$^{-13}$ & 0.95           & 2.2$\times$10$^{54}$ & 5.1$\times$10$^{-14}$ & 0.268          & 1.05 (=54.5/52) \\
6NW    & 400$\pm$50  &~~~6.2    & 0.112$\pm$0.015 & 1.8$\times$10$^{55}$ & 1.0$\times$10$^{-13}$ & $\dots$        & $\dots$              & $\dots$              &  $\dots$       & 1.42 (=50.0/35) \\
6SW    & 210$\pm$32  &~~~6.2    & 0.12$\pm$0.05   & 8.3$\times$10$^{54}$ & 5.7$\times$10$^{-14}$ & 0.95           & 9.7$\times$10$^{53}$ & 1.5$\times$10$^{-14}$ & 0.270          & 1.69 (=27.0/16) \\
\hline
W99$^{\mathrm{c}}$ & 4560& ~~~35 & 0.129           & 1.0$\times$10$^{56}$ & 6.5$\times$10$^{-12}$ & 2.4               & 4.3$\times$10$^{55}$ & 1.2$\times$10$^{-12}$ & 0.185 & 20 (=40/2) \\
C03$^{\mathrm{d}}$ & $\dots$& ~~~11 & 0.094$\pm$0.009 & 8.2$\times$10$^{56}$ & 7.2$\times$10$^{-12}$ & 0.7$^{+1.5}_{-0.5}$ & 5.1$\times$10$^{54}$ & 1.1$\times$10$^{-13}$ & 0.015  & 1.02 \\
\hline
\end{tabular}
\begin{list}{}{}
\item{$^{\mathrm{a}}$EM = $\int n_{\rm e}^2 dV$.}
\item{$^{\mathrm{b}}$Observed (absorbed) fluxes for the
    two-temperature models components in the energy range
    0.3-1.3~keV.}
\item{$^{\mathrm{c}}$\citet{1999A&A...343..599W}.}
\item{$^{\mathrm{d}}$\citet{2003ApJ...599.1189C}.}
\end{list}
\label{tab:spec_fits}
\end{table*}

The feature at $\sim 0.4$~keV can be identified with the 0.43~keV
\ion{N}{6} triplet, while the fainter feature at $\sim 0.6$~keV can be
associated with the 0.57~keV \ion{O}{7} triplet.  The occurrence of
spectral lines is suggestive of optically thin plasma emission,
confirming previous X-ray spectral analyses of S\,308
\citep{1999A&A...343..599W,2003ApJ...599.1189C}.  The predominance of
emission from the He-like species of nitrogen and oxygen over their
corresponding H-like species implies a moderate ionization stage of
the plasma. Furthermore, the relative intensity of the \ion{N}{6} and
\ion{O}{7} lines suggests nitrogen enrichment, since the intensity of
the \ion{O}{7} lines from a plasma with solar abundances would be
brighter than that of the \ion{N}{6} lines.

In accordance with their spectral properties and previous spectral
fits of the NW regions of S\,308 \citep{2003ApJ...599.1189C}, all the
X-ray spectra of S\,308 have been fit with XSPEC v12.7.0
\citep{Arnaud1996} for an absorbed two-temperature APEC optically thin
plasma emission model. The absorption model uses
\citet{1992ApJ...400..699B} cross-sections. A low temperature
component is used to model the bulk of the X-ray emission, while a
high temperature component is used to model the faint emission above
0.7~keV.  We have adopted the same chemical abundances as
\citet{2003ApJ...599.1189C}, i.e., C, N, O, Ne, Mg, and Fe to be 0.1,
1.6, 0.13, 0.22, 0.13, and 0.13 times their solar values
\citep{1989GeCoA..53..197A}, respectively, which correspond to the
nebular abundances.  The simulated two-temperature APEC model spectra
were then absorbed by an interstellar absorption column and convolved
with the EPIC-pn response matrices.  The resulting spectra were then
compared to the observed spectrum in the 0.3--1.3~keV energy range and
$\chi^{2}$ statistics are used to determine the best-fit models. A
minimum of 50 counts per bin was required for the spectral fit. The
foreground absorption ($N_\mathrm{H}$), plasma temperatures ($kT_1$,
$kT_2$) with 1-$\sigma$ uncertainties, and emission measures (EM$_1$,
EM$_2$) of the best-fit models are listed in
Table~\ref{tab:spec_fits}.  The best-fit models are overplotted on the
background-subtracted spectra in Figure~\ref{fig:spec_diffuse},
together with the residuals of the fits. Multi-temperature models do
not provide a substantial reduction of the value of the reduced
$\chi^{2}$ of the fit. We note that the values of the reduced
$\chi^{2}$ differ the most from unity for large regions, implying
inconsistencies of the relative calibrations across the FoV, but the
spectral fits still provide a fair description of the observed
spectrum. In the following sections we discuss the spectral fits of
the emission from the different morphological components of S\,308 in
more detail.
 
Spectral fits using models with varying chemical abundances of C, N,
and O were also attempted, but they did not provide any statistical
improvement of the fit.  In particular, models with N/O abundance
ratios different from those of the nebula resulted in notably worst
quality spectral fits.  As noted by other authors
\citep[see][]{2003ApJ...599.1189C,2011ApJ...728..135Z}, an
X-ray-emitting plasma with chemical abundances similar to those of the
optical nebulae seems at this moment to be the most adequate model for
the soft X-ray emission from WR bubbles.


\subsubsection{Properties of the Global X-ray Emission from S\,308}
\label{sec:globalprops}
The best-fit to the combined spectrum of the whole nebula results in
unphysically high values of the hydrogen absorption column density,
$N_\mathrm{H}$, well above the range $[0.2 -
1.05]\times10^{21}$\,cm$^{-2}$ implied by the optical extinction
values derived from Balmer decrement of the nebula
\citep{1992A&A...259..629E}. The effects of $N_\mathrm{H}$ and
nitrogen abundance on the $\chi^{2}$ of the spectral fits appear to be
correlated, i.e., models with high $N_\mathrm{H}$ and low nitrogen
abundance fit the spectra equally well as models with low
$N_\mathrm{H}$ and high nitrogen abundance.  If we adopt the high
absorption column density from the best-fit model
($N_{\mathrm{H}}\gtrsim3\times10^{21}$~cm$^{-2}$), the elevated
nitrogen abundance reported by \citet{2003ApJ...599.1189C} will not be
reproduced.  As the high absorption column density is not supported by
the optical extinction, in the subsequent spectral fits we will adopt
a fixed $N_\mathrm{H}$ of $6.2\times10^{20}$\,cm$^{-2}$ that is
consistent with the optical extinction measurements.  We note that
this choice results in an imperfect modeling of the spectral features
in the 0.3--0.5~keV range, as indicated by the S-shaped distribution
of residuals in this spectral region in Fig.~\ref{fig:spec_diffuse}.
If we allow the value of $N_{\mathrm{H}}$ to float during the spectral
fit, the improvement of the value of the reduced $\chi^{2}$ is
negible.

The parameters of the best-fit model, listed in the first line of
Table~\ref{tab:spec_fits}, show two plasma components at temperatures
$\sim$1.1$\times$10$^6$~K and $\sim$1.3$\times$10$^7$~K with an
observed flux ratio, $f_2/f_1\sim$0.09, corresponding to an intrinsic
flux ratio $F_2/F_1\sim$0.06. The total observed flux is
$\sim3\times10^{-12}$~erg~cm$^{-2}$~s$^{-1}$.  The intrinsic
luminosity at a distance of 1.5~kpc\footnote{See discussion about the
  distance to WR\,6 of \citet{2003ApJ...599.1189C}.}, after accounting
for a fraction of $\sim1/3$ of S\,308 which is not included in the
source apertures considered here, is
$\sim2\times$10$^{33}$~erg~s$^{-1}$.  The emission measure of the
best-fit to the combined spectrum, along with the spatial distribution
of the X-ray-emitting gas in a spherical thick shell with a thickness
$\sim$8\arcmin\ and inner radius of $\sim$11\arcmin, implies an
average electron density $n_e\sim 0.1$~cm$^{-3}$. We note that the
quality of the spectral fit is not exceptionally good, but more
sophisticated fits using multi-temperature models failed to improve
the quality of the fit. The proposed 2-T model, providing a fair
description of the spectral shape, should be considered as a first
approximate of the hot gas content and its physical conditions.
\\
\\

\subsubsection{Northwest Blowout (Region $\#2$)}

The northwest blowout of S\,308 has the brightest X-ray emission, with
a surface brightness
$\sim2.0\times10^{-18}$\,erg\,cm$^{-2}$\,s$^{-1}$\,arcsec$^{-2}$ and
its individual spectrum has a high signal-to-noise ratio.  The
spectral shape is consistent with those of the shell spectra, with a
prominent 0.43~keV \ion{N}{6} line, a weaker \ion{O}{7} line, and a
clear detection of X-ray emission to energies of 0.8--1.0~keV.  The
best-fit parameters are rather similar to those of the spectrum of the
entire nebula, with a marginally lower temperature for the hard
component.  We will adopt this value of the hard component temperature
for those regions whose spectra do not have an adequate count number
to fit this parameter.
 
\subsubsection{The Limb-Brightened Shell} 
\label{sec:shell}

The diffuse X-ray emission from S\,308 has a clear limb-brightened
morphology surrounding a cavity of diminished X-ray surface
brightness.  The emission from regions at the rim of this shell
(1NE, 1NW, 1SE, and 1SW) is relatively bright, with an averaged
surface brightness of the rim $\sim$
$1.2\times10^{-18}$\,erg\,cm$^{-2}$\,s$^{-1}$\,arcsec$^{-2}$.  All
individual spectra of the rim regions show the bright 0.43~keV \ion{N}{6}
emission line and indications of the weaker 0.57~keV \ion{O}{7}
emission line.  The hard component is faint, except for the
spectrum of region 1SW.  The fit to the combined spectrum confirms the
temperature of the soft component, but it is not possible to provide
statistical proof of the detection of the hard component.  The fits to
the individual spectra only provide upper limits for this component,
except for region 1SW where it seems relatively bright.

\subsubsection{The Central Cavity}

The level of X-ray emission from the innermost regions of the optical
shell of S\,308 is lower than that of its edge, with an averaged
surface brightness of
$5\times10^{-19}$\,erg\,cm$^{-2}$\,s$^{-1}$\,arcsec$^{-2}$, i.e.,
$\sim 2.5$--4.0 times fainter than the shell and blowout regions.  The
combined X-ray spectrum of the interior regions shown in
Figure~\ref{fig:spec_diffuse} indicates a stronger relative
contribution from the hard component. This is indeed confirmed by the
spectral fit: on average, the hard X-ray component has a flux
$\sim13\%$ that of the soft component.  There is a noticeable lack of
emission from this component in the region 6NW, but otherwise the
average contributions derived from the individual fits are higher than
for the spectra of apertures on the shell rim.

\subsubsection{Comparison with Previous X-ray Studies}

Table~\ref{tab:spec_fits} also lists the best-fit parameters of the
spectral fits to the diffuse X-ray emission from S\,308 obtained by
\citet{1999A&A...343..599W} and \citet{2003ApJ...599.1189C}.  It is
worthwhile discussing some of the differences with these previous
X-ray analyses.  The \citet{2003ApJ...599.1189C} joint fit of our
regions 1NW, 2NW, 3NW and 6NW yields very similar results to the ones
shown in Table~\ref{tab:spec_fits}.
For the second thermal component, the derived temperatures from our
spectral fits and those of \citet{2003ApJ...599.1189C} are consistent
with each other, but \citet{1999A&A...343..599W} provides a much
higher temperature for this component.  This discrepancy highlights
the difficulty of fitting the hard component using \emph{ROSAT} PSPC
data given its low spectral resolution, as well as the very likely
contamination of the \emph{ROSAT} PSPC spectrum of S\,308 by
unresolved hard point sources superposed on the diffuse emission.

\section{The Central Star WR6 (HD\,50896)} 
\begin{figure}[t]
\includegraphics[bb=18 200 592 700,width=1.0\linewidth]{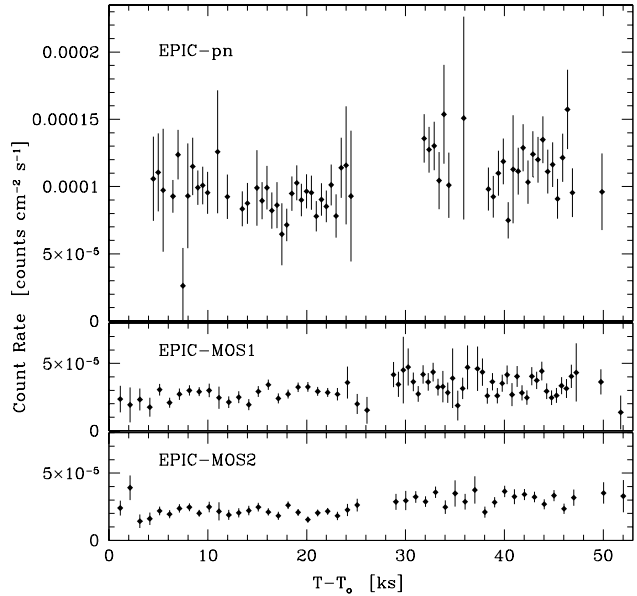}  
\vspace{0.1cm}
\caption{
EPIC-pn (top), MOS1 (center), and MOS2 (bottom) exposure-map-corrected 
light curves of WR\,6 in the 0.2--9.0~keV energy band.  
The time is referred to the starting time of the NE observation, 
2004-03-15T06:45:41 UTC.  
}
\label{fig:lc}
\end{figure}

\begin{figure*}[!htbp]
\includegraphics[width=1.0\linewidth]{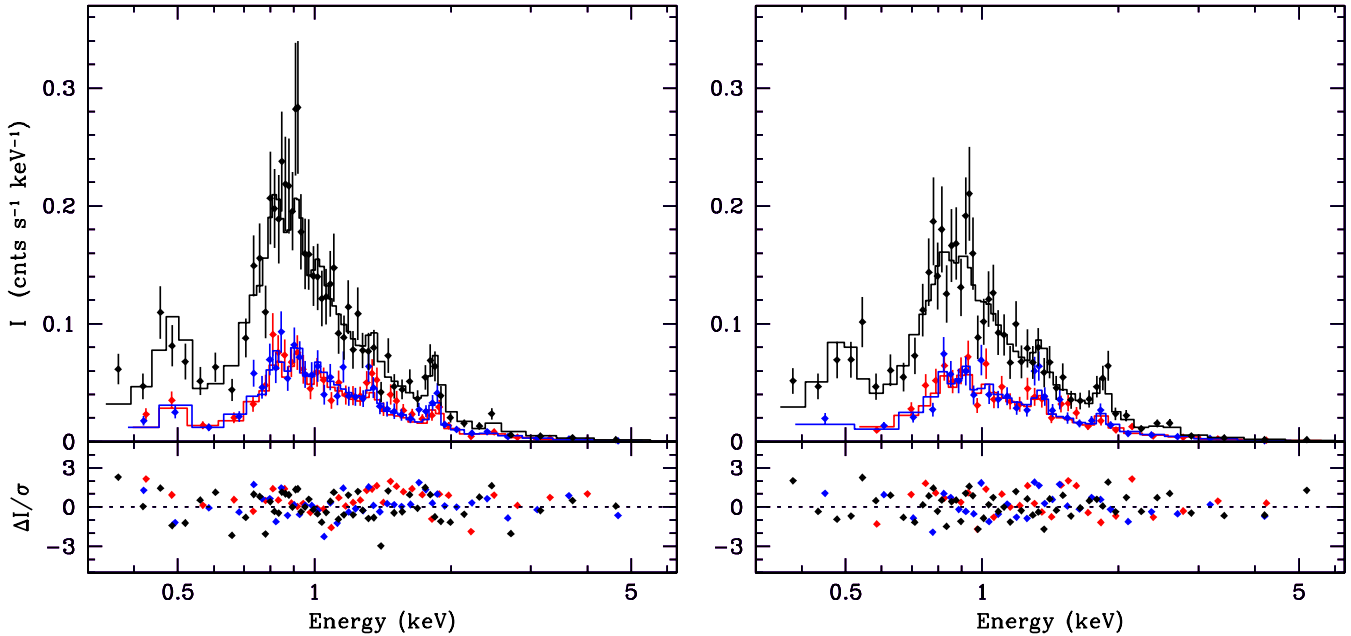}
\caption{
Background-subtracted \textit{XMM-Newton} EPIC-pn (black), MOS1 (red), 
and MOS2 (blue) spectra of WR\,6 obtained during the observation of 
the NE (\textit{left}) and SW (\textit{right}) pointings of S\,308.  
}
\label{fig:wr_spectrum}
\end{figure*}

\begin{table*}[ht!]
\caption{Spectral Fits of HD\,50896}
\centering
\begin{tabular}{lrrr}
\hline\hline\noalign{\smallskip}
\multicolumn{1}{c}{Parameter}          &
\multicolumn{1}{c}{NE Spectrum}        &
\multicolumn{1}{c}{SW Spectrum}        &
\multicolumn{1}{c}{\citet{2002ApJ...579..764S}} \\
\hline\noalign{\smallskip}
$N_{H}$ [$\times$10$^{21}$\,cm$^{-2}$]   &  6.4$^{+1.0}_{-0.9}$   &   5.9$^{+1.2}_{-0.9}$        & 4.0$^{+0.4}_{-0.6}$ \\
$kT_1$ [keV]                            &  0.28$^{+0.03}_{-0.04}$   &   0.28                & 0.6$^{+0.4}_{-0.4}$ \\
$A_1$ [cm$^{-5}$]                       & 7.1$\times$10$^{-3}$  &   5.9$\times$10$^{-3}$    & 3.4$\times$10$^{-5}$ \\
$kT_2$ [keV]                            &  2.5$^{+1.5}_{-0.5}$   &   2.48                   & 3.5$^{+0.7}_{-0.5}$\\
$A_2$ [cm$^{-5}$]                        & 1.8$\times$10$^{-4}$ &   2.2$\times$10$^{-4}$    & 1.0$\times$10$^{-5}$\\
$\chi^2$/DoF                            & 1.10=137.0/124      &  1.10=113.2/103           & 1.08=234.5/217  \\
$f_{1}$ (0.2--10 keV)   [$\times$10$^{-12}$ erg\,cm$^{-2}$\,s$^{-1}$] &  1.14               &  1.23          &  0.97 \\
$f_{1}$ (2.5--10 keV)   [$\times$10$^{-12}$ erg\,cm$^{-2}$\,s$^{-1}$] &  0.33               &  0.39        &  0.31 \\
$f_{2}$ (0.2--10 keV)   [$\times$10$^{-12}$ erg\,cm$^{-2}$\,s$^{-1}$] &  0.58                 &  0.70 &  0.49 \\
$F_{1}$ (0.2--10 keV)   [$\times$10$^{-12}$ erg\,cm$^{-2}$\,s$^{-1}$] & 19.8                  & 17.6 &  2.90 \\ 
$F_{1}$ (2.5--10 keV)   [$\times$10$^{-12}$ erg\,cm$^{-2}$\,s$^{-1}$] &  0.35                 &  0.41 &  0.33 \\
$F_{2}$ (0.2--10 keV)   [$\times$10$^{-12}$ erg\,cm$^{-2}$\,s$^{-1}$] &  1.10                 &  1.35 &  0.78 \\
\hline
\end{tabular}
\label{tab:WR}
\end{table*}

The central star of S\,308 is WR\,6 (a.k.a.\ HD\,50896), which has a
spectral subtype WN4 \citep{1988A&A...199..217V}.  The star is
detected by the \textit{XMM-Newton} EPIC cameras in the NE and SW
pointings of the nebula.  The EPIC-pn, EPIC-MOS1, and EPIC-MOS2 count
rates are 160$\pm$4 counts~ks$^{-1}$, 65$\pm$2 counts~ks$^{-1}$, and
65$\pm$2 counts~ks$^{-1}$, respectively, from the NE observation, and
141$\pm$4 counts~ks$^{-1}$, 53$\pm$2 counts~ks$^{-1}$, and 52$\pm$2
counts~ks$^{-1}$, respectively, from the SW observation using a source
aperture of $50''$ in radius. These count rates appear to imply that
the X-ray luminosity of WR\,6 diminished by 10--20\% from the NE to
the SW observations, which were only $\sim8$~hours apart\footnote{ The
  on-axis \textit{XMM-Newton} images presented by
  \citet{2002ApJ...579..764S} revealed a faint X-ray source at a
  radial distance of $\sim$57\arcsec\ from WR\,6.  The flux from this
  source is $(2\pm1)$\% that of WR\,6 and thus it is not likely that
  fraction of the emission from this source entering into the aperture
  used for WR\,6 would contribute significantly to the observed X-ray
  variability.}.  We note, however, that these count rates are largely
affected by vignetting due to the offset position of WR\,6 on the EPIC
cameras.  Indeed, the light curves shown in Figure~\ref{fig:lc}, after
accounting for the effects of vignetting, may imply the opposite,
i.e., that the X-ray flux of WR\,6 was slightly higher in the second
(SW) observation than in the first (NE) observation.

In Figure~\ref{fig:wr_spectrum} we present the EPIC
background-subtracted spectra for the two different epochs.  Following
\citet{2002ApJ...579..764S}, we have modeled these spectra with a
two-temperature VAPEC model with initial abundances set to those shown
in Table~1 of \citet{1986A&A...168..111V}.  The fit allowed the
foreground absorption column density, temperatures, and abundances of
N, Ne, Mg, Si, and Fe to vary \citep{2002ApJ...579..764S}. Table 3
displays the parameters resulting from our best-fit models: absorption
column densities $N_{\rm H}$, plasma temperatures $T$, normalization
parameters $A$\footnote{$A=1\times10^{-14} \int n_{e}^{2}dV/4 \pi
  d^{2}$, where $d$ is the distance, $n_{e}$ is the electron density,
  and $V$ the volume in cgs units.}, observed (absorbed) fluxes $f$,
and intrinsic (unabsorbed) fluxes $F$. Model fits for the spectra from
the NE and SW observations are listed separately alongside those from
\citet{2002ApJ...579..764S} for comparison. The column density and
temperatures of the two components are within 1-$\sigma$ of one
another among the three different models.  The observed fluxes are
also consistent, although the \citet{2002ApJ...579..764S} flux seems
to be a bit lower, while our fluxes are closer to the ones derived
from the October 1995 \textit{ASCA} observations of WR\,6
\citep{1996AAS...189.7717S}. The total observed fluxes and the
observed fluxes of the hot thermal component, $f_{2}$, listed in
Table~\ref{tab:WR} may indicate a hardening of the X-ray emission from
WR\,6 during the last observation. To assess this issue, we performed
statistical evaluation of the lightcurves showed in
Figure~\ref{fig:lc} using the HEASOFT task \textit{lcstats} and found
no significant variations. Thus, WR\,6 does not show evidence of
variability in time-scales of hours.

We would like to point out that the absorption column density obtained
from our best fits are in good agreement with the values obtained from
\citet{2002ApJ...579..764S}, which are higher than that used to fit
the soft X-ray emission from the nebula. Such higher column density
values are commonly observed towards massive stars such as WR stars
and are recognized to be caused by absorption at the base of the wind
\citep{1981ApJ...250..677C,1994ApJ...436L..95C,2009A&A...506.1055N,2010AJ....139..825S,2011A&A...527A..66G}.
\\
\\

\section{Discussion}
\label{sec:discussion}

The \textit{XMM-Newton} images and spectra analyzed in the previous
sections reveal that the hot plasma in S\,308 is spatially distributed
in a thick shell plus the northwest blowout, with most emission being
attributable to a hot plasma at $\sim$1.1$\times$10$^6$ K. For an
adiabatically shocked stellar wind, its temperature is determined by
the stellar wind velocity, $kT = 3\mu m_{\mathrm{H}} V_{w}^{2}/16$,
where $\mu$ is the mean particle mass for fully ionized gas
\citep{Dyson1997}. Therefore, the temperature expected for the shocked
stellar wind of WR\,6, with a terminal wind velocity of
1700~km~s$^{-1}$ \citep{1998A&A...333..251H} and $\mu\gtrsim1.3$
\citep[][]{1986A&A...168..111V}, would be $T > 8\times10^{7}$~K, in
sharp contrast with the observed temperature. The same issue has been
pointed out for the WR bubble NGC\,6888 by several authors
\citep[see][and references therein]{2011ApJ...728..135Z}, and it is
also a common issue in planetary nebulae \citep[e.g.,
NGC\,6543;][]{2001ApJ...553L..69C}.

Electron thermal conduction has been proposed as a mechanism capable
of reducing the temperature of the hot plasma in shocked stellar wind
bubbles. Thermal conduction was applied by \citet{1977ApJ...218..377W}
to stellar wind bubbles to produce a self-similar solution for the
density and temperature structure in bubbles. The soft X-ray
luminosities predicted by Weaver et al.'s model for the Omega Nebula
and the Rosette Nebula, according to the stellar wind parameters of
their associated young clusters (M\,17 and NGC\,2244, respectively),
are several orders of magnitude higher than those observed
\citep{2003ApJ...593..874T,2003ApJ...590..306D}. Thus, the standard
Weaver et al.\ model for a stellar wind bubble with thermal conduction
cannot be taken at face value. Recent work by
\citet{2008A&A...489..173S} in the context of planetary nebulae, which
are produced in a very similar manner to WR wind bubbles, has
calculated the time-dependent radiation-hydrodynamic evolution of
planetary nebula wind bubbles including thermal conduction in 1D
models with spherical symmetry. In these models, the cold shell
material from the previous AGB superwind phase is evaporated into the
hot bubble. Saturated conduction was taken into account in these
calculations by limiting the electron mean free path and it was found
that thermal conduction was able to lower the temperature and raise
the density at the edge of the hot bubble enough to explain the soft
X-ray emission and low X-ray luminosities observed in some planetary
nebulae \citep[e.g.,][]{2001ApJ...553L..69C}.

In the case of WR bubbles, \citet{2011ApJ...737..100T} presented
time-dependent 2D radiation-hydrodynamic models of the evolution of
the CSM around single massive stars, including classical and saturated
thermal conduction. They found that in the absence of a magnetic
field, thermal conduction does not seem to play an important role in
shaping WR bubbles, but that models with thermal conduction have
slightly greater soft-band luminosities than those without thermal
conduction. They suggested that the morphology of S\,308 could result
from a star with initial mass of 40~$M_{\odot}$ whose stellar
evolution model includes stellar rotation
\citep{2003A&A...404..975M}. They obtained that $\sim$20,000~yr after
the onset of the WR phase, the X-ray-emitting gas will present a
clump-like morphology with an electron density of
$n_{e}\sim0.1$~cm$^{-3}$ inside an optical ($T\sim10^{4}$~K) shell
with radius of $\sim9$~pc.

Whereas the \citet{2011ApJ...737..100T} models reproduce the
morphology and X-ray luminosity of wind-blown bubbles, their
simulations predict higher temperatures of the hot plasma that result
in X-ray spectra that do not match the observed spectral shape. This
might imply that additional physical processes must be taken into
account. An interesting alternative to thermal conduction for the
apparently low ionization state of the plasma is non-equilibrium
ionization (NEI).  \citet{2010ApJ...718..583S} calculate the
timescales to reach collisional ionization equilibrium (CIE) for
ionized plasmas and their results suggest that, for values of the
parameters relevant to S\,308 (derived electron density and time in
the WR stage; \citealp{2011ApJ...737..100T}), the CIE assumptions may
not hold. These ideas will be pursued in subsequent works.

\section{Summary and conclusions}

We present \textit{XMM-Newton} observations of three fields of the WR
bubble S\,308 which, in conjunction with the observation of its NW
quadrant presented by \citet{2003ApJ...599.1189C}, map most of the
nebula except for its southernmost section.  We have used these
observations to study the spatial distribution of the X-ray-emitting
material within this bubble, to derive global values for its physical
conditions ($T_{e}$, $n_{e}$), and to search for
their spatial variations among different morphological components of
the nebula.

The X-ray emission from S\,308 is found to have a limb-brightened
morphology, with a shell thickness 5\arcmin--8\arcmin, and extend to
the northwest blowout region. The X-ray-emitting shell is notably
larger along the SE-NW direction than along the SW-NE direction, and
it is always confined by the optical shell of ionized material. The
X-ray surface brightness decreases notably from the blowout region and
the western rim shell to the shell interior, where the X-ray emission
falls to background levels. The western quadrants are also brighter
than the eastern quadrants.

The X-ray emission from S\,308 shows prominent emission from the
He-like triplet of \ion{N}{6} at 0.43~keV and fainter emission of the
\ion{O}{7} 0.57~keV triplet, and declines towards high energies, with
a faint tail up to 1~keV.  This spectrum can be described by a
two-temperature optically thin plasma emission model with temperatures
$\sim 1.1 \times 10^6$~K and $\sim 1.3 \times 10^7$~K. The latter
component is notably fainter than the former by at least a factor of
$\sim 6$. There is an appreciable difference in the relative
contributions of the hot component to the X-ray-emitting gas between
the rim and the nebula interior, of which the latter has a higher
contribution from the hard component. The total X-ray luminosity is
estimated to be $\sim2\times 10^{33}$~erg~s$^{-1}$ for a distance of
1.5~kpc.

\acknowledgements This research was supported by the NASA
\textit{XMM-Newton} Guest Observer Program Grant NNG\,04GH99G. SJA and
JAT acknowledge financial support from DGAPA-UNAM through grant PAPIIT
IN100309. JAT also thanks CONACyT, CONACyT-SNI (Mexico) and CSIC
JAE-PREDOC (Spain) for a student grant.  JAT and MAG are partially
funded by grant AYA2001-29754-C03-02 of the Spanish Ministerio de
Econom\'{i}a y Competitividad.


\end{document}